\documentclass[structabstract]{aa}  
\usepackage{graphicx}
\usepackage{txfonts}
\begin{document}
   \title{A stringent upper limit to 18cm radio emission from the extrasolar planet system $\tau$ Bo\"{o}tis}

\titlerunning{Upper limit to radio emission from the exoplanet system $\tau$ Bo\"{o}tis}

\author{A. Stroe, I.A.G. Snellen \and H.J.A. R\"{o}ttgering}

\institute{Leiden Observatory, Leiden University, Postbus 9513, 2300 RA, Leiden, The Netherlands}

\date{}

\abstract{It has been speculated for many years that some extrasolar planets may
emit strong cyclotron emission at low radio frequencies in the range
10-100 MHz. Despite several attempts no such emission has yet been seen.}
{The hot Jupiter system $\tau$ Bo\"{o}tis is one of the nearest (d=15
pc) exoplanets known to date.  The gravitational influence of this massive hot Jupiter (M=8 M$_{\rm{jup}}$) has locked the star-planet system, making the star rotate in P$\sim$3.3 days, similar to the orbital period of the planet. From the well established correlation between stellar rotation and radio luminosity, it is conceivable that the $\tau$ Bo\"otis system emits strong radio emission at significantly higher frequencies than currently probed, which we aimed to investigate with this work.}{We observed $\tau$ Bo\"otis with the Westerbork Synthesis Radio Telescope (WSRT) at a frequency of 1.7 GHz. for 12 hours in spectral line mode, reaching a noise level of 42 $\mu$Jy/beam at the position of the target.}{No 18cm radio emission is detected from $\tau$ Bo\"otis, resulting in a 3$\sigma$ upper limit of 0.13 mJy, corresponding to a 18cm radio luminosity of $<3.7\times10^{13}$ erg s$^{-1}$ Hz$^{-1}$. We observe $\tau$ Bo\"otis to be two orders of magnitude fainter than expected from the stellar relation between radio luminosity and rotation velocity. }{This implies that either the $\tau$ Bo\"otis system is underluminous in the radio compared to similar fast-rotating stars, or that we happened to observe the target during a low state of radio emission.}

\keywords{Extrasolar planets}

\maketitle

\section{Introduction}

It has been speculated for many years that some extrasolar planets may
emit strong cyclotron emission at low radio frequencies in the range
10-100 MHz (Winglee et al. 1988; Zarka et al. 1997; Farrell et al. 1999; Bastian et al. 2000; Zarka et al. 2001; Lazio et al. 2004; Stevens 2005; Griessmeier et al. 2005; Zarka 2006, 2007). 
During magnetic storms, our own Jupiter can outshine the sun
at these frequencies by several orders of magnitude. These storms are
caused by the interaction of the magnetic fields of Jupiter, Io, and
charged particles in the solar wind - causing the cyclotron
radiation. It is expected that for those exoplanets that orbit their
parent star at very short orbital distances (hot Jupiters, a$\sim$0.05 AU),
this interaction is so intense that their low frequency radio emission can
be seen with current-day radio telescopes. Despite several
attempts no such emission has yet been seen, making it an appealing
science objective for the new low-frequency radio telescope LOFAR (Farrel et al. 2004).

$\tau$ Bo\"otis is one of the nearest (d=15 pc) exoplanets known to date. The 
F7V star (M = 1.3 M$_{\rm{Sun}}$, R=1.331 R$_{\rm{Sun}}$) is orbited by a hot Jupiter with $m \sin i$ = 3.9 M$_{\rm{Jup}}$ with a period of P=3.31 days (Butler et al., 1997). Spectropolarimetric observations of the host star reveals differential rotation with periods between 3.0 and 3.7 days from the equator to the poles. It implies that the planet is synchronized with the star's rotation at intermediate latitudes (Catala et al. 2006). The chromospherically active star is also found to produce X-rays at a luminosity of $6.8\times10^{28}$ erg s$^{-1}$ (Kashyap, Drake, \& Saar 2008). Recently, Brogi et al. (2012) have shown that the planet orbits the star at an inclination of 44.5$\pm$1.5$^o$, by detecting the signature of orbital motion in the spectrum of the planet.

Lazio \& Farell (2007) observed $\tau$ Bo\"otis at 74 MHz ($\lambda$=4.0 m) with the Very Large Array, resulting in an upper limit of 150 mJy, constraining the low-frequency cyclotron radiation from the system. The scaling law from Zarka et al. (2001) predicts a 74 Mhz flux density of 1 Jy for this system. However, it is expected that fast-rotating stellar systems should also produce radio emission at higher frequencies. e.g. Stewart et al. (1988) present a relation between 8.4 GHz ($\lambda$=3.6 cm) peak luminosities and the rotational velocity of 51 late-type F, G, and K stars - including both single stars (giants, sub-giants and dwarf stars) and active components of close binary-systems, 
\begin{equation}
\log_{10} [ L / (R/R_{\rm{Sun}})^{2.5}] = 12.0 + (2.5 \pm0.5) \log_{10}v
\end{equation}
where $L$ is the peak radio luminosity of the star (in erg s$^{-1}$ Hz$^{-1}$), $R$ is the stellar radius, and $v$ is the rotational velocity (in km s$^{-1}$). For $\tau$ Bo\"otis this would imply a peak luminosity in the 3$-$60 mJy range. To our knowledge the star has never been observed at these high frequencies, and only the 1.4 GHz ($\lambda$=21 cm) NVSS survey (Condon et al. 1998) provides a 5$\sigma$ upper limit of 2.5 mJy. 
We therefore observed $\tau$ Bo\"otis with the Westerbork Synthesis Radio Telescope (WSRT) at 18cm targeting its conceivable high-frequency emission. Section 2 describes the observations and data reduction, and section 3 the results and discussion.

\section{Observations and data reduction}

\begin{table} [t!]
\caption{Details of the WSRT observations}          
\label{tab:radioobs}      
\centering          
\begin{tabular}{l l}     % 8 columns 
\hline\hline       
Source       & $\tau$ Bo\"otis          \\       
RA (J2000)   & 13 47 15.74340 *  \\                     
DEC (J2000)  & +17 27 24.8552 * \\
Configuration & Maxi-short \\
Date & Sept 03, 2011 \\
Duration (hours) & 12.0 \\
$\nu$ range (MHz)   & 1640 - 1786\\           
Number of IFs       & 8 \\            
Channels per IF     & 128 \\            
Bandwidth (MHz)   & 20 \\
Channel width (kHz)  & 156.25 \\
Polarisations & XX, YY \\
Beam size  & $61.9'' \times 14.5''$ \\
Weighting & robust \\
Theoretical rms noise & 20 $\mu$Jy beam$^{-1}$ \\ \hline
\multicolumn{2}{l}{\scriptsize * van Leeuwen 2007.}\\
\end{tabular}
\end{table}
$\tau$ Bo\"otis was observed for 12 hours in spectral line mode in the L band (18 cm) with the Westerbork Synthesis Radio Telescope (WSRT) on September 3, 2011. The maxi-short configuration with an integration time per visibility point of 60 sec was used. Two polarisations (XX and YY) and eight spectral windows (IFs) of 20 MHz bandwidth and 128 channels each, were recorded. The resulting frequency coverage ranged from 1640 MHz to 1786 MHz. \\
The data set was reduced using the Common Astronomy Software Applications (CASA). It was flagged automatically for autocorrelations and shadowing and then manually for Radio Frequency Interference (RFI). Antenna RT5 was used as testbed for a new receiver, did not record interferometric data and was thus flagged. Two calibrators were also observed for 30 min each: 3C147 before and CTD93 after the observation of $\tau$ Bo\"otis. Their flux densities were set on the Perley-Butler (2010) scale to 19.43 Jy and 4.41 Jy at 1665 MHz (18 cm), respectively. The data was corrected for bandpass and gain variations, flagging channels 0 to 4 and 101 to 128 because of the roll-off of the bandpass. The data for the science target was split and underwent several rounds of phase only self-calibration and one round of amplitude and phase self-calibration. \\
The sidelobes from two bright galaxies affected region centred on the target source and were removed using the peeling technique (Noordam 2004). The brightest source could not be removed completely most probably due to an imperfect model that does not capture the extended emission correctly. The image was produced using robust weighting set to 0.0 (Briggs 1995) and was cleaned manually using clean boxes,  yielding a synthesized beam of $61.9'' \times 14.5''$ . Taking into consideration the 50$\%$ data lost from flagging, the theoretical rms was 20 $\mu$Jy beam$^{-1}$. The observed rms noise $\sigma_{obs}$ is composed of two contributions: the thermal noise, $\sigma_{th}$, caused mainly by the telescope electronics and galactic background, and the confusion noise, $\sigma_{conf}$, that mostly affects the central part of the field of view. Considering the two contributions to be uncorrelated:
\begin{equation}
\sigma_{obs} = \sqrt{\sigma_{th}^2 + \sigma_{conf}^2} \;.
\end{equation}
The measured rms noise at the edges of the image in areas free of emission was 32 $\mu$Jy beam$^{-1}$, whereas in the centre it reaches 42 $\mu$Jy beam$^{-1}$. A confusion noise of roughly 27 $\mu$Jy beam$^{-1}$ added in quadrature to the noise at the edges of the field can account for the higher rms value at the centre. \\
The image was corrected for the primary beam attenuation, leading to an rms value within a 4 arcmin square region centred on the position of $\tau$ Bo\"otis of 42.6 $\mu$Jy beam$^{-1}$ with a maximum of 131.8 $\mu$Jy beam$^{-1}$. Therefore, we can pose a 5 sigma upper of 0.21 mJy$^{-1}$ limit on the radio flux of $\tau$ Bo\"otis at 18 cm.
\begin{figure*}[h!]
\centering
\includegraphics[width=14.0cm]{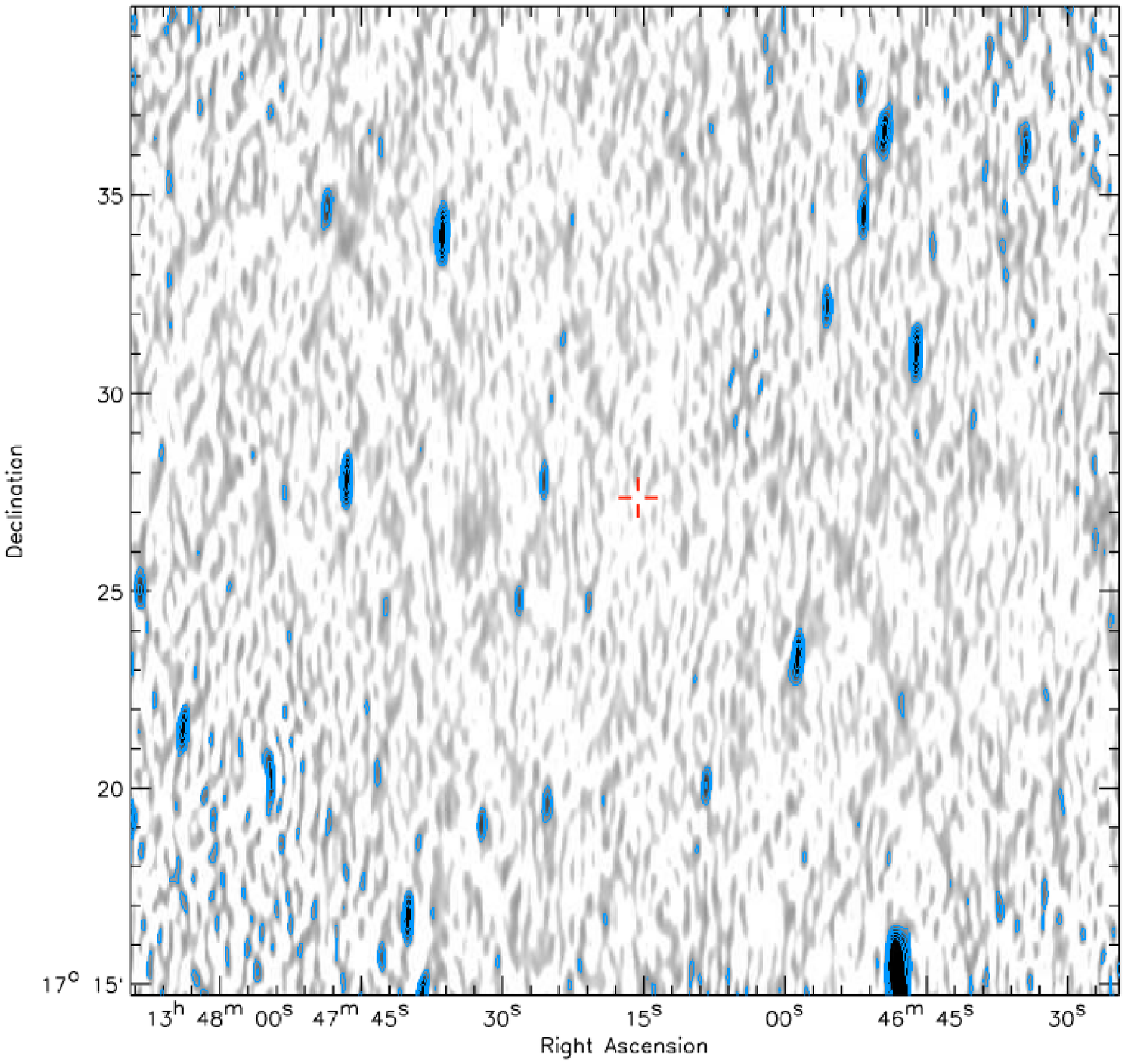}
\caption{WSRT 1640 - 1786 MHz primary beam corrected image of the $\tau$ Bo\"otis field. Contours at $[1,2,4,8,16]\times 4\sigma_\mathrm{rms}$ are overlaid.  The location of $\tau$ Bo\"otis is marked in red. The beam size is $61.9'' \times 14.5''$. }
\label{fig:TAU_BOO}
\end{figure*}

\section{Results and Discussion}
The resulting image is shown in Figure 1, with the position of $\tau$ Bo\"otis marked by the cross. The system is not detected, resulting in a 3$\sigma$ upper limit of 0.13 mJy. This indicates that its 18cm radio luminosity, at $<$3.7x10$^{13}$ erg Hz$^{-1}$ sec$^{-1}$ during the observations was significantly lower than the stellar peak luminosity found of stellar systems with similar equatorial rotation velocities. The radius and rotational period of $\tau$ Bo\"otis indicate a rotational velocity at the equator of $\sim$20 km sec$^{-1}$.  From the stellar rotation - radio luminosity relation of Stewart et al. (1988) at 8.4 GHz, as given in Eq. 1, $\tau$ Bo\"otis is expected to to have a peak 8.4 GHz luminosity of 3.8x10$^{15}$ erg sec$^{-1}$ Hz$^{-1}$, corresponding to a flux density of 13 mJy (between 3 and 60 mJy when taking into account the uncertainty of the relation). Slee et al. (1987) and references therein indicate that these stars have generally flat spectra with F$_{\nu} \propto \nu^{0\pm0.3}$, meaning that at 18 cm (1.6 GHz) the expected flux should be more than 60\% of that at 8.4 GHz, corresponding to $>$2$-$40 mJy. The 3$\sigma$ upper limit is therefore a factor 15$-$300 lower than peak fluxes for similar fast-rotating stars. 

These observations imply that either the $\tau$ Bo\"otis system is underluminous in the radio compared to similar fast-rotating stars, or that we happen to observe the target during a low state of radio emission. The latter is in line with flaring nature of the emission during which the radio luminosity can very by more than an order of magnitude on time scales of days to weeks (e.g. Slee et al. 1987). Note that 8 stars in the Stewart et al. (1988) sample have meaningful upperlimits, implying that they were not observed during a high state of activity. Taking into account that all stars were observed typically ten times, it implies that they are on average flaring for $\sim$20\% of the time. A more intense monitoring program is required to show this for $\tau$ Bo\"otis.

That $\tau$ Bo\"otis is underluminous at radio frequencies is less clear if the well known correlation between the radio and X-ray luminosities of magnetically active stars (e.g. G\"udel 2002) is considered. According to this relation, the X-ray luminosity of $\tau$ Bo\"otis of $6.8\times10^{28}$ erg sec$^{-1}$ (Kashyap, Drake, \& Saar 2008) points to a radio luminosity of 10$^{13-14}$ erg sec$^{-1}$ Hz$^{-1}$, corresponding to an expected flux density at about the level of the 3$\sigma$ upper limit of our observations. Note that it may appear that, since Jupiter exhibit cyclotron bursts in narrow, low-frequency bands, and similar behaviour is expected for hot Jupiters, this could explain the absence of emission from tau Bootis.  However, in this study we probed the broad-band high-frequency component expected to come from the host star, not from the planet.
 
\begin{acknowledgements}
We thank the WSRT team, in particular Gyula Jozsa, for help with the observations.
\end{acknowledgements}

\end{document}